\documentclass[aps,prl,showpacs,nobibnotes,twocolumn,preprintnumbers,
nobalancelastpage,superscriptaddress]{revtex4}
\usepackage{latexsym}
\usepackage{dcolumn}
\usepackage{bm}
\usepackage{amsmath}
\usepackage{natbib}
\input{epsf}
\newcommand{\be}{\begin{equation}}
\newcommand{\ee}{\end{equation}}
\newcommand{\ba}{\begin{eqnarray}}
\newcommand{\ea}{\end{eqnarray}}
\newcommand{\bi}{\begin{itemize}}
\newcommand{\ei}{\end{itemize}}
\newcommand{\bfi}{\begin{figure}[!t]
\epsfxsize=9cm
\epsffile}
\newcommand{\efi}{\end{figure}}
\newcommand{\efig}{\end{figure*}}

\newcommand{\la}{\lesssim}
\newcommand{\ga}{\gtrsim}
\newcommand{\mpch}{{\rm Mpc}/h}
\newcommand{\hmpc}{h/{\rm Mpc}}

\begin{document}
\title{Determination of the large scale volume weighted halo velocity bias in simulations}
\author{Yi Zheng}
\affiliation{Key Laboratory for Research in Galaxies and Cosmology, Shanghai
  Astronomical Observatory, 80 Nandan Road, Shanghai, 200030,
  China}
\affiliation{Korea Astronomy and Space Science Institute, Daejeon 305-348, Republic of
Korea}
\author{Pengjie Zhang}
\email[Email me at: ]{zhangpj@sjtu.edu.cn}
\affiliation{Center for Astronomy and Astrophysics, Department of
  Physics and Astronomy, Shanghai Jiao Tong
  University, 955 Jianchuan road, Shanghai, 200240}
\affiliation{IFSA Collaborative Innovation Center, Shanghai Jiao Tong
University, Shanghai 200240, China}
\affiliation{Key Laboratory for Research in Galaxies and Cosmology, Shanghai
  Astronomical Observatory, 80 Nandan Road, Shanghai, 200030,
  China}
\author{Yipeng Jing}
\email[Email me at: ]{ypjing@sjtu.edu.cn}
\affiliation{Center for Astronomy and Astrophysics, Department of
  Physics and Astronomy, Shanghai Jiao Tong
  University, 955 Jianchuan road, Shanghai, 200240}
\affiliation{IFSA Collaborative Innovation Center, Shanghai Jiao Tong
University, Shanghai 200240, China}

\begin{abstract}
A profound assumption in peculiar velocity cosmology is $b_v=1$  at sufficiently large scales, where $b_v$ is the volume weighted halo(galaxy) velocity bias with respect to the matter velocity field. However, this fundamental assumption has not been robustly verified in numerical simulations. Furthermore,  it is challenged by structure formation theory (BBKS, 1986, ApJ; Desjacques and Sheth, 2010, PRD), which predicts the existence of velocity bias (at least for proto-halos) due to the fact that halos reside in special regions (local density peaks).  The major obstacle to measure the volume weighted velocity from N-body simulations is an unphysical sampling artifact. It is entangled in the measured velocity statistics and becomes significant for sparse populations. With recently improved understanding of the sampling artifact (Zhang, Zheng and Jing, 2015, PRD; Zheng, Zhang and Jing, 2015, PRD), for the first time we are able to {\it appropriately correct this sampling artifact and then robustly measure the volume weighted halo velocity bias}. (1) We verify  $b_v=1$ within $2\%$ model uncertainty at $k\la 0.1\hmpc$ and $z=0$-$2$ for halos of mass $\sim 10^{12}$-$10^{13} h^{-1} M_\odot$, and, therefore, consolidates a foundation of the peculiar velocity cosmology. (2) We also find statistically significant signs of $b_v\neq 1$ at $k\ga 0.1\hmpc$. Unfortunately, whether this is real or caused by residual sampling artifact requires further investigation. Nevertheless, cosmology based on $k\ga 0.1\hmpc$ velocity data shall be careful this potential velocity bias. 
\end{abstract}
\maketitle


\section{Introduction}
Large scale peculiar velocity is maturing as a powerful probe of
cosmology. Peculiar velocity  directly responds to the gravitational pull of all clustered matter and energy, making it a precious tool to study dark matter (DM), dark energy, and the nature of gravity (e.g. \cite{Peacock01,Zhang07c,Guzzo08,Jain08,Li13}). 
Measuring peculiar velocity at cosmological distances with the conventional method of distance indicators is challenging, albeit improving (e.g., \cite{Johnson14,Watkins14}). Alternatively, redshift space distortion (RSD) provides a way of measuring peculiar velocity at cosmological distances,  free of the  otherwise overwhelming contamination of Hubble flow.  It enables  $\sim 1\%$ accuracy in the velocity power spectrum measurement at $z\sim 1$ (e.g. Fig. 2.3, \cite{BigBOSS}), through stage IV dark energy surveys such as DESI and Euclid. 

A profound assumption in cosmology based on peculiar velocity is that the velocity bias $b_v$ of galaxies vanishes at large scales ($b_v=1$),  namely, that the galaxy velocity field is statistically identical to that of the matter velocity field at large scales. The  strong equivalence principle predicts that galaxies sense the same acceleration as ambient DM particles. Hence, one would naturally expect statistically identical velocity for galaxies and DM particles,  at $\ga 10 \mpch$ scales where the only operating force is gravity.  However, a loop hole in this argument is that galaxies and their host halos only reside in special regions (local density peaks). The same  environmental difference is known to cause $b_v<1$ in proto-halos \cite{BBKS,DesjacquesI,DesjacquesII,Elia12,Baldauf14,Biagetti14}. However, due to the stochastic relation between proto-halos and real halos \cite{Colberg00}, it is non-trivial to extrapolate this prediction to real halos where galaxies reside. Since $v\propto fDb_v$ at large scale, uncertainties in $b_v$ lead to systematic error in all existing  $fD$ measurements \cite{Chuang13},
\be
\frac{\delta (fD)}{fD}=1-b_v^{-1}\ . 
\ee
Here $f\equiv d\ln D/d\ln a$ and $D$ is the linear density growth factor. Therefore we have to understand $b_v$ to $1\%$ or better to make the peculiar velocity competitive with other dark energy probes. 

A key intermediate step to understand the galaxy velocity bias is to understand the halo velocity bias \footnote{Galaxies in a halo have extra velocities relative to the host halo. The correlation length of this velocity field is $\la 1\mpch$, so it does not contribute to velocity at $k\sim 0.1\hmpc$ of interest. Therefore the large scale halo velocity and galaxy velocity are identical, statistically speaking. }.  N-body simulations are ideal to  robustly clarify this issue.   What is most relevant for cosmology, in particular RSD cosmology, is the {\it volume weighted} halo velocity bias at large scales \footnote{For example, the velocity power spectrum determined from RSD is volume weighted in the framework of \cite{paperI}.  }. Unfortunately, measuring the volume weighted velocity statistics through inhomogeneously and sparsely distributed particles/halos  is highly challenging, due to a sampling artifact \cite{DTFE96,Pueblas09,paperII,Zhang14,Zheng14a,Jennings14}.

 This sampling artifact arises from the fact that we only have information of velocities at positions of particles/halos. Therefore the sampling of the volume weighted velocity field is incomplete. Even worse, since the particle/halo velocity field is correlated with the particle/halo distribution, the sampling of volume weighted velocity field is imperfect. Such completeness and imperfection leads to inaccurate measurement of velocity statistics, which we call the ``sampling artifact". For  sparse populations,  it can cause $\sim 10\%$ systematic underestimation of the velocity power spectrum at $k=0.1\hmpc$ \cite{Zhang14,Zheng14a,Jennings14}. Even worse, it also depends on the intrinsic LSS (large scale structure) fluctuation in the particle distribution and its correlation with velocity \cite{Zheng14a}.  This sampling artifact is by itself unphysical, in the sense that it solely arises from the limitation of robustly measuring the volume weighted velocity statistics {\it given} the inhomogeneously and sparsely distributed velocity data.  Given its existence, the rawly measured velocity bias from simulation is a mixture of the real velocity bias and the sampling artifact in the following form:
\ba
\label{eqn:SA}
\hat{b}_v({\rm wrong})=b_v({\rm true})\times {\rm sampling \ artifact}\ . 
\ea
Namely, the raw bias measurement $\hat{b}_v$ is wrong by a multiplicative factor caused by the sampling artifact. Without rigorous correction, the sampling artifact can be misinterpreted as a significant velocity bias and mislead the peculiar velocity cosmology. 

 Therefore, robustly understanding the sampling artifact is a prerequisite for reliably  measuring the true velocity bias.  For this purpose,  we developed the theory of the sampling artifact  in \cite{Zhang14} and rigorously confirmed the existence of the sampling artifact in simulations \cite{Zheng14a}. We further tested the theory against simulations and improved it to $1\%$ accuracy at $k=0.1\hmpc$ for populations with number density $\ga 10^{-3}(\mpch)^{-3}$\cite{Zheng14a}. In particular, \cite{Zheng14a} demonstrates the sharp distinction between a real velocity bias and the sampling artifact, for DM samples.  It first constructs samples with a fraction of the simulation DM particles randomly selected from all the simulation particles. By construction, the velocity statistics of the random samples shall be statistically identical to those of the sample including all simulation DM particles.  Namely $b_v({\rm DM})=1$.  However, the raw measurement shows $\hat{b}_v({\rm DM})\neq 1$ of high significance \cite{Zheng14a}. The fake $\hat{b}_v({\rm DM})\neq 1$ then clearly demonstrates the sampling artifact (Eq. \ref{eqn:SA}). 

In the current paper, we applied this improved understanding of the sampling artifact \cite{Zhang14,Zheng14a} to robustly eliminate it  in velocity measurement and correctly determine the  true {\it volume weighted halo velocity bias} for the first time. This differs from existing numerical works on measuring velocity bias \cite{Colin00,Elia12,Baldauf14,Jennings14}, which either focus on proto-halos,  the density weighted halo velocity statistics, or the volume weighted halo velocity mixed with the sampling artifact. 

\section{Simulation specifications}
 We analyze the same J1200 N-body simulation in \cite{Zheng14a}, run with a particle-particle-particle-mesh (${\rm P^3M}$) code \cite{Jing07}. It adopts a $\Lambda$CDM cosmology with  $\Omega_m=0.268$, $\Omega_\Lambda=0.732$, $\Omega_b=0.045$,  $\sigma_8=0.85$,  $n_s=1$ and  $h=0.71$.  It has $1200{\rm Mpc}/h$ box size, $1024^3$ particles and mass resolution of $1.2\times 10^{11} M_\odot/h$. The halo catalogue is constructed by Friends-of-Friends (FOF) method with a linking length $b=0.2$. Gravitationally unbound ``halos'' have been excluded from the catalogue. In total we have $N_h=6.18\times10^6$ halos  with at least 10 simulation particles, at $z=0$. We choose the mass weighted center as the halo center and the velocity averaged over all member particles as the halo velocity. We try three mass bins detailed in table \ref{Table:massbin}. 

\begin{table}[htb]
\begin{center}
\begin{tabular}{r|c|c|c|c|ccccc}\hline
Set ID & mass range & $\langle M\rangle$ & $N_h/10^5$ &$n_h$ &$b_h({\rm density})$\\\hline
\textit{A1}($z=0.0$) & $10$-$3700$ & $39$ & $7.5$ & $4.4$&$1.3$\\
               $z=0.5$\ \ &  $10$-$2300$ & 30 & $5.9$ & 3.5& 1.8 \\
               $z=1.0$\ \ & $10$-$950$   & 24 & $3.7$ & 2.1& 2.6\\
               $z=2.0$\ \ & $10$-$400$   & 19 & $1.3$ &$0.76$  &4.3\\ \hline\hline
\textit{A2}($z=0.0$) & $1.2$-$10$ & $2.8$  & $54$ &$32$ &$0.8$ \\
                 $z=0.5$\ \  &  $1.2$-$10$   & 2.8  & $52$ & 31 &1.1 \\
                 $z=1.0$\ \  &  $1.2$-$10 $  & 2.7  & $46$ &  27 & 1.5 \\
                 $z=2.0$\ \  &  $1.2$-$10$   & 2.5  & $31$ & $18$ &2.4\\\hline\hline
\textit{B}($z=0.0$)& $2.3$-$3700$ & $13$　& $31$ & $18$&$1.0$  \\\hline
\end{tabular}
\end{center}
\caption{Three sets of halo mass bins.  The mass unit is $10^{12}M_\odot/h$ and the halo number density $n_h$ has unit of $10^{-4}(\mpch)^{-3}$. $\langle M\rangle$ is the mean  halo mass. {\bf $N_h$ is the total halo number in one halo mass bin.} The density bias $b_h$ is averaged around $k=0.01\hmpc$. The mass bin $B$ at $z=0$ has density bias of unity, designed for better control of the sampling artifact. }
\label{Table:massbin}

\end{table}

\section{Correcting the sampling artifact}
We aim to measure  the halo  velocity bias defined in Fourier space,
\be
\label{eqn:bv}
b_v(k)\equiv \sqrt{\frac{P^v_{h,E}(k)}{P^v_{{\rm DM}, E}(k)}}\ .
\ee
The subscript ``E'' denotes the gradient (irrotational) part of the velocity, which is most relevant for peculiar velocity cosmology. The subscripts``$h$'' and ``DM''  refer to halos and DM simulation particles respectively. Throughout this paper, we restrict ourself to  the {\it volume weighted} power spectrum. We adopt the NP (Nearest Particle) method \cite{paperII} to sample the velocity field on $256^3$ uniform grids.  Before correcting the sampling artifact, the measured velocity power spectrum $\hat{P}_E(k)$ differs from its true value by a factor $C(k)\equiv \hat{P}^v_E(k)/P^v_E(k)$.
We found that \cite{Zheng14a} 
\ba
\label{eqn:CT}
 C(k)&\simeq& \langle e^{i{\bf k}\cdot{\bf D}}\rangle^2 e^{k^2\xi_D(r=\alpha/k)/3}\equiv C_T(k)\ .
\ea
Here, ${\bf D}$ is the deflection field pointing from a particle used for velocity assignment to the corresponding grid point to which the velocity is assigned. Reference \cite{Zhang14} showed that ${\bf D}$ fully captures the sampling artifact. $\xi_D$ is the spatial correlation of ${\bf D}$.  For $\alpha=1/2$ and $\bar{n}_P\sim 10^{-3} (\mpch)^{-3}$, $C_T(k)$ agrees with the actual $C(k)$ to $\sim 1\%$ at $k\leq 0.1\hmpc$ \cite{Zheng14a}.  The subscript ``T'' denotes that it is our theory prediction. We caution that the theoretically predicted $C_T(k)$ may deviate from the true $C(k)$ since the theory prediction is not exact. Furthermore, $C$ (or $C_T$) of dark matter particles can differ from that of dark matter halos.  

\bfi{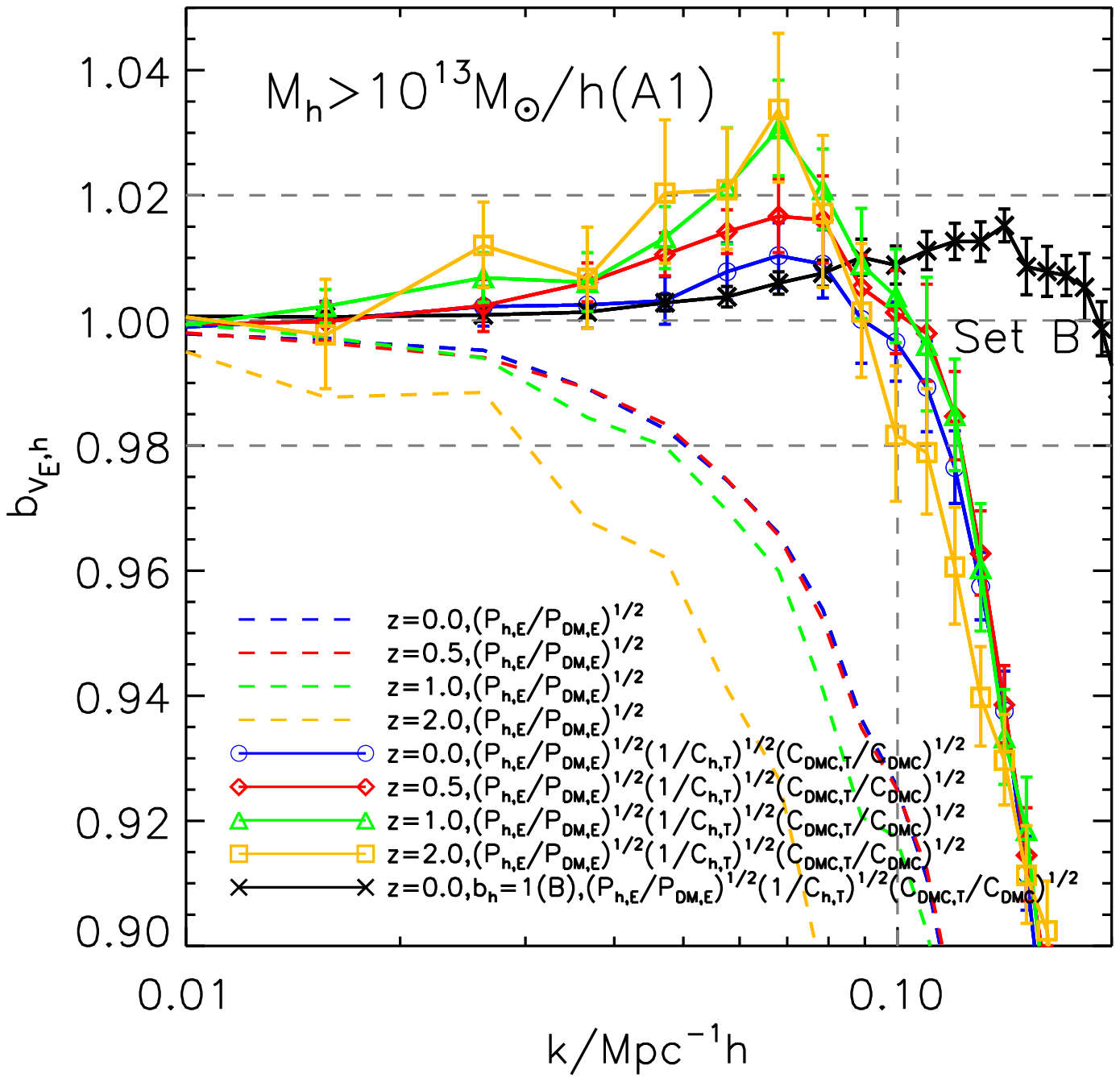}
\caption{The measured velocity bias for mass bin A1 and B at different redshifts (Table 1). The data points connected with solid lines are the final results, {\it after correcting the sampling artifact}. The error bars are the r.m.s dispersions between  $10$ realizations of DM control samples.  For comparison, we also show the raw measurements (dashed curves), which are essentially the sampling artifact, unrelated to the physical velocity bias. Our correction of the sampling artifact has percent level model uncertainty at $k=0.1\hmpc$ and we have highlighted it with the two dashed straight lines with somewhat arbitrary value $1\pm 0.02$.  After correction, we find $b_v=1$ at $k\la 0.1\hmpc$ within $2\%$ model uncertainty and hence consolidate this fundamental assumption of peculiar velocity cosmology. At $k\ga 0.1\hmpc$, there are signs of $b_v\neq 1$, which require further rigorous investigation/verification.   Measuring the velocity bias to higher accuracy requires improvement over the existing understanding of the sampling artifact. \label{fig:bva}}
\efi
\bfi{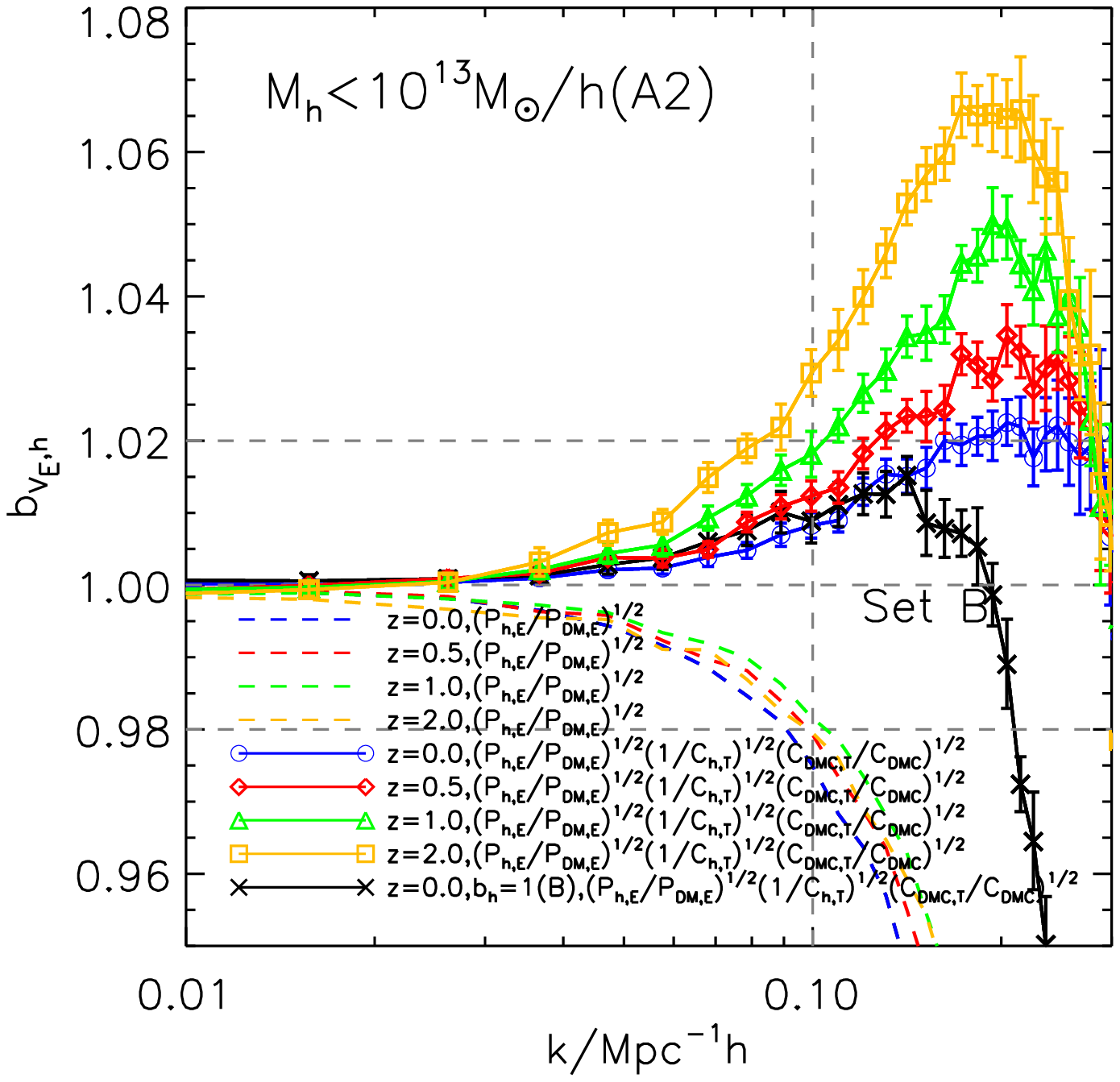}
\caption{Same as Fig. \ref{fig:bva}, but for mass bin A2 and B (Table 1). The apparent ``anti-bias'' before correction disappears after correction. Roughly speaking, the corrected velocity bias at $k<0.1\hmpc$ is consistent with unity within $2\%$ model uncertainty. Nevertheless, there are signs of increasing $b_v$ with increasing $k$ and $z$. \label{fig:bvb}}
\efi

 We take two steps to correct for the sampling artifact.  {\bf Step one}. We use Eq. \ref{eqn:CT} to  correct for the bulk of the sampling artifact. {\bf Step two}. There are residual sampling artifact since our theory is not perfect ($C_T\neq C$). We further correct this residual sampling artifact with the aid of DM control samples (DMCs). They are constructed by randomly selecting simulation DM particles from the full simulation sample with the requirement $\sigma_D({\rm DMC})=\sigma_D({\rm halo})$ \footnote{For this requirement, DMCs are more sparse than the halo samples.}.  $\sigma_D\equiv \langle {\bf D}^2\rangle^{1/2}$ is the dominant factor determining the sampling artifact \cite{Zhang14,Zheng14a}. The halo sample and DMCs have identical sampling artifact at the $k\rightarrow 0$ limit and similar sampling artifact elsewhere.  Hence to greater accuracy than Eq. \ref{eqn:CT}, we expect $C_{h,T}/C_{h}\simeq C_{\rm DMC,T}/C_{\rm DMC}$.  The subscripts ``$h$''  and``DMC'' denote properties of halos and DM control samples, respectively.  We then obtain
\ba
\label{eqn:c2}
b_v(k)&\simeq&\sqrt{\frac{\hat{P}^v_{h,E}(k)}{P^v_{{\rm DM},E}(k)}} \sqrt{\frac{1}{C_{h,T}(k)}}\sqrt{\frac{C_{\rm DMC,T}(k)}{C_{\rm DMC}(k)}}\ .
\ea
The terms on the r.h.s. are, respectively, the raw velocity bias measurement without correcting the sampling artifact, the step one correction, and the step two correction. $P^v_{\rm DM,E}$ is measured from the full J1200 simulation sample, which is essentially free of sampling artifact due to its high $\bar{n}_P$ \cite{paperII}. All the correction terms ($C_{h,T}$, $C_{\rm DMC,T}$ and $C_{\rm DMC}$) are directly calculated from the J1200 simulation.

The inaccuracy of Eq. \ref{eqn:c2} increases with $k$. For peculiar velocity cosmology to be competitive, at least we shall utilize measurement at $k\leq 0.1\hmpc$. So we choose $k=0.1\hmpc$ as the pivot scale for quoting the accuracy.  Overall we expect $\sim 1\%$ accuracy \footnote{For all the investigated halo bins, except $A1(z=2)$,  $\bar{n}_h>2\times 10^{-4} (\mpch)^{-3}$. Based on \cite{Zheng14a}, we expect $\sim 1\%$ accuracy in  the step one correction (Eq. \ref{eqn:CT}). For $A1(z=2)$, the number density is lower ($0.76\times 10^{-5}$). However Eq. \ref{eqn:CT} is more accurate at $z=2$ than $z=0$ \cite{Zheng14a}. Hence we also expect Eq. \ref{eqn:c2} to be accurate at $\sim 1\%$ level. The step two correction can further improve the accuracy.}, extrapolating from the DM cases. We caution the readers on this $\sim 1\%$ uncertainty in the measured $b_v(k)$ (Figs. \ref{fig:bva} \& \ref{fig:bvb}).  Somewhat arbitrary, we quote the systematic error in the measured $b_v$ as $2\%$ at $k<0.1\hmpc$. Therefore, only if  $|b_v-1|>0.02$ at $k<0.1\hmpc$, are we capable of detecting a non-unity velocity bias. More accurate measurement of velocity bias requires better correction of the sampling artifact to below $1\%$, either by improved modelling of the sampling artifact, or by improved velocity assignment method (e.g. \cite{yukriging}).

\bfi{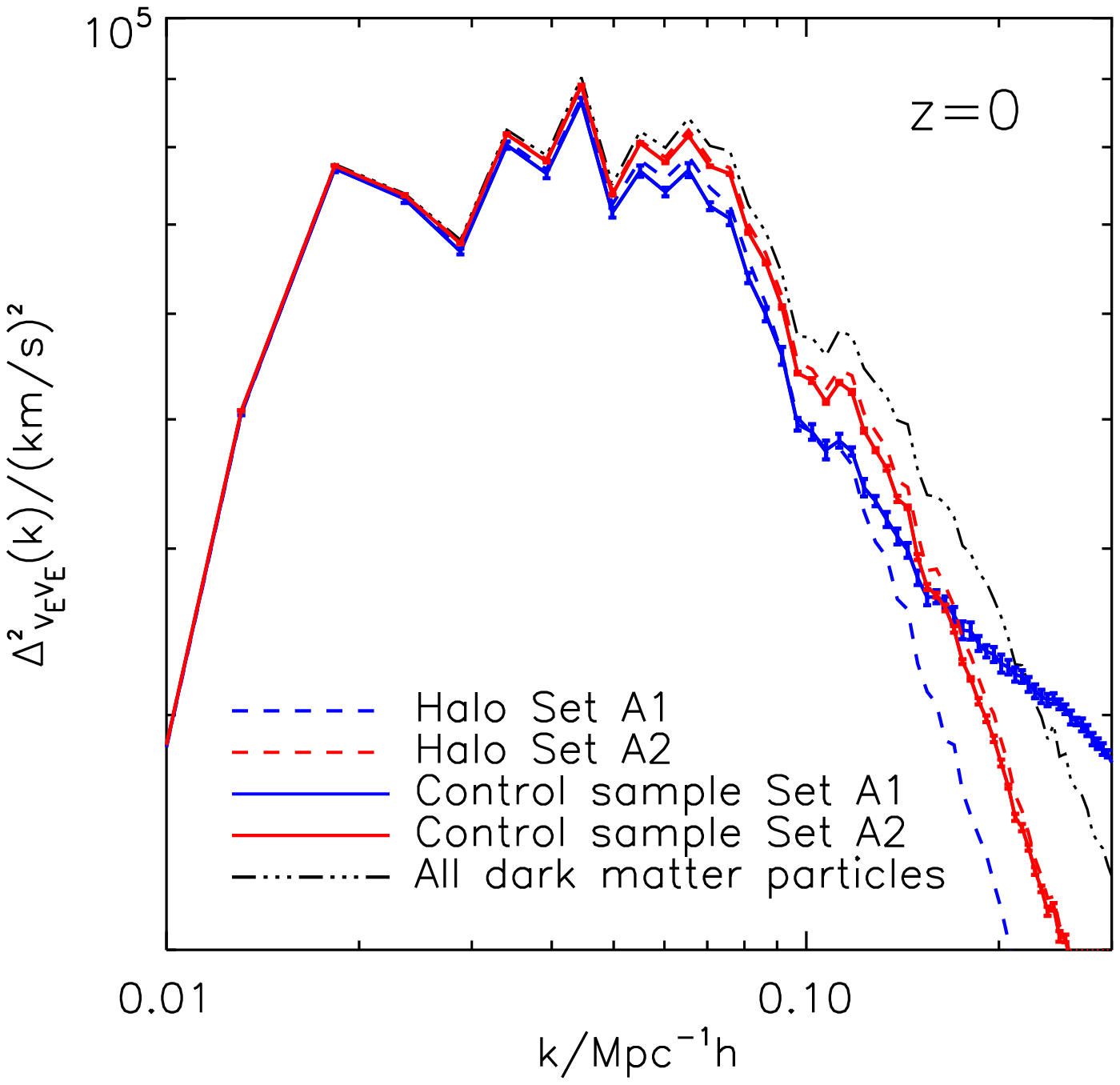}
\caption{The sampling artifact in the velocity power spectrum measured in N-body simulations, which causes systematic underestimation at $k\la 0.1\hmpc$.   (1) The halo velocity power spectra (dash lines) are lower than measured from all DM simulations particles (dash-dot line). (2) The velocity power spectra of DM control samples containing a fraction of all simulation particles are also lower. Member particles in the control samples are randomly selected from the full simulation particles and hence must have statistically identical velocity power spectra.  Therefore the observed deficit in the DM velocity power spectrum is caused by the sampling artifact \cite{Zheng14a}. (3) When the number density of DM control samples and halo samples are identical, they have similar (but not identical) velocity power spectra and similar deficit with respect to the full DM sample. These are solid evidences of significant sampling artifact in the measured halo velocity power spectrum. The most crucial step in measuring the halo velocity bias is to understand and correct this sampling artifact. This is the sole purpose of our two preceding works \cite{Zhang14,Zheng14a}, which show that the sampling artifact depends on not only the number density, but also on the intrinsic LSS fluctuation and its correlation with the velocity field.  \label{fig:sa}}
\efi
\section{No velocity bias at $k\la 0.1\hmpc$}
 Figs. \ref{fig:bva} \& \ref{fig:bvb}  show $b_v$ for all mass bins listed in Table 1. The raw measurements suggest an ``anti-bias'', unanimous for all mass bins at all redshifts. This is most significant for the more massive bin $A1$, reaching $b_v\sim 0.9$ at $k=0.1\hmpc$ (Fig. \ref{fig:bva}). Possibly by coincidence, this ``anti-bias'' agrees well with theoretical predictions of proto-halos based on linear/Gaussian statistics. However, we have solid evidences  that it is essentially an illusion caused by the sampling artifact. The sampling artifact causes systematic suppression of $P^v$  \cite{Zhang14,Zheng14a,Jennings14}, mimicking an anti-bias.  In another word, the apparent ``anti-bias'' is {\it unreal} in the sense that is unrelated to the true velocity statistics of halos and is irrelevant for cosmology. 

Theoretically, we expect the sampling artifact to exist for  any populations of inhomogeneously distributed objects, and its impact to be significant for sparse populations \cite{Zhang14}. It has been robustly detected for the case of DM simulation particles \cite{paperII,Zheng14a}. Therefore, it must also exist for DM halos \cite{Zhang14}. Fig. \ref{fig:sa} further consolidates this theoretical prediction. It shows that the DM control samples containing a fraction of DM simulation particles have smaller $P^v$ than the full DM sample at $k\la 0.1\hmpc$.  Furthermore, $P^v$ decreases with decreasing number density. If  the number densities of these DM control samples match those of halo samples, their $P^v$ match each other closely, especially at $k<0.1\hmpc$. This behavior holds for all three mass bins and four redshifts investigated. The DM control samples are constructed by randomly selecting DM simulation particles, so by construction the difference in $P^v$ to the full DM sample should not exist and any difference must be caused by the sampling artifact.  The similarities between DM control samples and halo samples then strongly suggest that the ``anti-bias'' implied by the raw measurement is merely the sampling artifact and  is therefore  {\it unrealistic}.  The bin $A1$ with $M>10^{13}M_\odot/h$ is a factor of $\sim 10$ more sparse than $A2$ with $M<10^{13}M_\odot/h$, so it suffers from a larger sampling artifact, $\sim 10\%$ at $k=0.1\hmpc$. 

Hence, it is essential to correct for the sampling artifact. After applying the two step corrections (Eq. \ref{eqn:c2}),  the ``anti-bias'' disappears and we find $b_v(k\leq 0.1\hmpc)=1$ within $2\sigma$ statistical uncertainty, for the A1 bin at all redshifts.   Taking the extra $2\%$ systematic uncertainty into account, we find no evidence on a non-unity velocity bias at $k\leq 0.1\hmpc$ for halos bigger than $10^{13} M_\odot/h$. At $k\leq 0.1\hmpc$, $b_v$ of bin $A2$ after correcting the sampling artifact shows statistically significant evidence for $b_v>1$, opposite to the ``anti bias'' that raw measurement suggests. However, once the $2\%$ systematic uncertainty is taken into account, again we find no evidence for $b_v\neq 1$ at $k\leq 0.1\hmpc$ (perhaps except $z=2$).

How solid are these results? To check it, we construct a mass bin $B$ with $M>2.3\times 10^{12}M_\odot/h$. It has identical large scale LSS fluctuation as DMCs, so we can better handle its sampling artifact by comparing with DMCs.  Thus we treat the $b_v$ measurement of bin $B$ as the most accurate halo velocity bias measurement that we can achieve. Again we find $b_v=1$ at $k\leq 0.1\hmpc$.  Therefore we conclude that $b_v=1$ at $k<0.1\hmpc$ within $2\%$ model uncertainty.  Settling the issue of whether $b_v=1$ at greater accuracy requires further improvement over existing understanding of the sampling artifact \cite{Zheng14a} or better velocity assignment method.

The vanishing velocity bias ($b_v=1$) within $2\%$ model uncertainty at $k<0.1h/$Mpc verifies a fundamental assumption in peculiar velocity cosmology.  However, from the theoretical viewpoint, this result is quite surprising, as linear theory predicts $b_v(k=0.1\hmpc)\simeq 0.9$ for $\sim 10^{13}M_\odot/h$ proto-halos (peaks in initial/linearly evolved density field)\cite{BBKS,DesjacquesI,DesjacquesII,Baldauf14,Biagetti14}. The predicted $b_v<1$ arises from correlation between density gradient and velocity at initial density peaks. A number of processes may weaken/destroy this correlation and hence make the velocity bias disappear. First is the stochasticity in proto halo-halo relation. A fraction of halos today do not correspond to initial density peaks and a fraction of  initial density peaks do not evolve into halos today (e.g. \cite{Colberg00}). Second, halos move from their initial positions. They tend to move towards each other and, hence, modify their velocity correlation. Third, the density and velocity evolution has non-negligible nonlinearity (e.g. \cite{Colberg00,paperII}), and, hence, non-Gaussianity. This alters the predicted velocity bias based on Gaussian statistics.

\section{Velocity bias $b_v\neq 1$ at $k\ga 0.1\hmpc$?}
On the other hand,  at $k\ga 0.1\hmpc$ there are signs of $b_v\neq 1$ and signs of mass and redshift dependences. 
(1) For mass bin $A2$, the data suggest that $b_v>1$ and $b_v-1$ increases with increasing $k$ and $z$. The excess is statistically significant  at $z=2$ and $k\geq 0.1\hmpc$. (2) In contrast, bin $A1$ ($>10^{13}M_\odot/h$) has $b_v(k>0.1\hmpc)<1$ at the $1\%$ to $10\%$ level, which is also statistically significant. 

 Due to the opposite signs of $b_v-1$ for more and less massive halos, an even more significant behavior is that smaller halos seem to move faster at $k>0.1\hmpc$ and the difference reaches $10\%$ at $k\sim 0.15\hmpc$. If this difference is indeed intrinsic, instead of a residual sampling artifact, it could be caused by different environments where different halos reside. Small halos tend to live in filaments and have extra infall velocity with respect to large halos. The infall velocity has a correlation length of typical filament length of tens of Mpc, and, hence, shows up at  $k\ga 0.1\hmpc$.  Unfortunately, our understanding of the sampling artifact at $k>0.1\hmpc$ is considerably poorer \cite{Zheng14a}. Therefore we are not able to draw decisive conclusions, other than that cosmology based on the peculiar velocity at $k\ga 0.1\hmpc$ must keep caution on this potential velocity bias.

\section{Conclusions and discussions}
This paper presents  the first determination of  volume weighted halo velocity bias through N-body simulations. The raw measurements suffer from a severe sampling artifact which could be misinterpreted as a significant ``velocity bias.'' We are able to appropriately correct the sampling artifact  following our previous works \cite{Zhang14,Zheng14a} and measure the {\it physical} velocity bias. Two major findings are as follows:
\bi
\item $b_v=1$ at $k\leq 0.1\hmpc$ within $2\%$ model uncertainty. It consolidates the peculiar velocity cosmology;
\item Signs of $b_v\neq 1$ at $k\ga 0.1\hmpc$ and signs that $b_v-1$ depends on redshifts, scales and halo mass. Although we are not able to robustly rule out the possibility of a residual sampling artifact, it raises the alarm of using $k\ga 0.1\hmpc$ velocity data to constrain cosmology.
\ei

Accurate measurement of the velocity bias in simulations heavily relies on robust correction of the sampling artifact. The sampling artifact depends on not only the halo number density, but also on the intrinsic LSS fluctuation of the halo distribution and its correlation with the halo velocity field \cite{Zheng14a}. It is for this reason that  our  understanding of the sampling artifact at $k\sim 0.1\hmpc$ so far is no better than $1\%$. Therefore, we caution the readers that the measured velocity bias has $\sim 1\%$ (or somewhat arbitrarily $2\%$) model uncertainty (systematic error). For a similar reason, we cannot fully quantify the accuracy of Eq. \ref{eqn:c2} and the accuracy in the sampling artifact corrected $b_v$ \footnote{Another reason is that we do not have corresponding halo samples of identical mass and intrinsic LSS clustering, but with halo number density $\bar{n}_h\rightarrow \infty$ and hence free of sampling artifact. This differs from the case of DM, for which in principle we can increase the simulation mass resolution to reach this limit.  }. We know it is more accurate than Eq. \ref{eqn:CT} and have estimated its accuracy by extrapolating from the DM cases \cite{Zheng14a}. Nevertheless, this ambiguity prevents us from drawing an unambiguous conclusion on whether the found $b_v\neq 1$ at $k>0.1\hmpc$ is real, or whether $b_v$ deviates from unity by less than $1\%$ at $k<0.1\hmpc$. Therefore a major future work will be to improve understanding of the sampling artifact (e.g., discussions in the Appendix of \cite{Zheng14a}). Furthermore, we will explore other velocity assignment methods which may alleviate the problem of the sampling artifact.  We will also extend to galaxy velocity bias with mock galaxy catalogues, where the sampling artifact should also be corrected.

 Finally, we address that the RSD determines  velocity indirectly  through the galaxy number density distribution and, therefore, the RSD inferred velocity statistics can be free of the sampling artifact.  This is another advantage of RSD over conventional velocity measurement methods. It is only when comparing the RSD determined velocity power spectrum with that measured in simulations that we must worry about the sampling artifact in the simulation part. 

{\bf Acknowledgement}.---
We thank Cris Sabiu's for his help in proofreading this paper.
This work was supported by the National
Science Foundation of China 
(Grants No. 11025316, No. 11121062,  No. 11033006, No. 11320101002, and No. 11433001), the National
Basic Research Program of China (973 Program 2015CB857001), the
Strategic Priority Research Program ``The Emergence of Cosmological
Structures" of the Chinese Academy of Sciences (Grant
No. XDB09000000), and the key laboratory grant from the Office of Science and Technology, Shanghai Municipal Government (No. 11DZ2260700).

\bibliography{mybib}
\end{document}